\documentclass[prb,aps,epsf,twocolumn,floatfix,floats]{revtex4}
\usepackage{epsfig}
\newcommand{\be}{\begin{equation}}
\newcommand{\ee}{\end{equation}}
\newcommand{\bea}{\begin{eqnarray}}
\newcommand{\eea}{\end{eqnarray}}

\def\nn{\nonumber\\}

\begin{document}
%%%%%%%%%%%%%%%%%%%%%%%%%%%%%%%%%%%%%%%%%%%%%%%%%%%%%%%%%%%%%%%%%%%
\title{Finite-temperature lineshapes in gapped quantum spin chains}
%%%%%%%%%%%%%%%%%%%%%%%%%%%%%%%%%%%%%%%%%%%%%%%%%%%%%%%%%%%%%%%%%%%
\author{Fabian H.L. Essler$^{(a)}$ and Robert M. Konik$^{(b)}$}
\affiliation{
$^{(a)}$ The Rudolf Peierls Centre for Theoretical Physics, Oxford
University, Oxford OX1 3NP, UK\\ 
$^{(b)}$ CMPMS Department, Brookhaven National Laboratory, Upton
NY 11973, USA }
\begin{abstract}
We consider the finite-temperature dynamical structure factor (DSF) of
gapped quantum spin chains such as the spin one Heisenberg model
and the disordered transverse field Ising model.
At zero temperature the DSF in these models is dominated by a
delta-function line arising from the coherent propagation of
single particle modes. Using methods of integrable quantum field
theory we determine the evolution of the lineshape at low
temperatures.  We show that the line shape is in general asymmetric in energy
and we discuss the relevance of our results for the
analysis of inelastic neutron scattering experiments on gapped spin
chain systems such as ${\rm CsNiCl_3}$ and ${\rm YBaNiO_5}$.
\end{abstract}

\maketitle
Quasi one-dimensional spin chains are materials where
quantum fluctuations give rise to 
striking strongly correlated phenomena.
An exemplar of such behavior is the distinction, first identified by 
Haldane \cite{haldane} over 20 years ago, between integer 
and half-integer isotropic spin chains.  The former
are generically gapped while the latter are generically gapless.
This effect is topological in origin and arises from the presence of a quantized Berry's
phase in the effective model describing the chains. 
Contemporary examples of strong correlations in spin chains
revolve around the role the multi-excitation continuum play in their physics.
This continuum has been studied both experimentally\cite{igor,kenz1,kenz2} and theoretically
\cite{3p} and is understood to be the origin of 
a process known as spectrum termination, where
coherent excitations cross into a multi-excitation continuum and then
experience rapid decay.\cite{igor1}  

In order to probe the dynamical behavior of spin chains, 
inelastic neutron scattering is the premier tool.
In particular, using inelastic neutron scattering it is possible
to determine with impressive accuracy
the spectrum of spin excitations together with their
lifetimes.\cite{igor2}  The theoretical counterpart of such measurements is
a computation of the dynamical structure factor (DSF).  For theoretical
models that admit an integrable continuum field theoretic description,
the computation of the zero temperature
DSF is possible with impressive accuracy.\cite{smirnov,review}  However at finite temperatures, while
important progress has been made,\cite{muss,rmk,alt} a general theoretical framework
has yet to be settled upon.
It is one aim of this letter to outline a promising approach.

We do so against an experimental mise en sc\`ene of particular relevance.  
In a system that supports a coherent, gapped, magnetic single-particle
excitation at $T=0$, the question arises of how the corresponding delta-function in the
%DSF broadens at non zero-temperatures. \cite{young,damle,xu,kenz3,kenz4,kenz5}
DSF broadens at non zero-temperatures. \cite{young,damle,xu,kenz3,kenz4}
A partial answer to this question has been given by Sachdev and
collaborators:\cite{young,damle} they demonstrated that at 
temperatures far below the gap the broadening in the immediate vicinity of the $T=0$ gap is Lorentzian
in form. In the present work, using our approach to computing
finite temperature DSFs, we determine the {\it entire} lineshape.
As our central finding, we demonstrate that the lineshape
is always asymmetric in energy, a feature that becomes more pronounced
as the temperature increases.
While we focus upon the lineshape in gapped quantum spin chains, 
we stress that this approach is applicable to the computation of general
response functions in generic integrable continuum models, such as those considered in Ref.(\onlinecite{alt,damle1}).

\noindent{\bf General Theoretical Framework:}
The systems we study all have representations as general
Heisenberg models:
\begin{equation}\label{ei}
H = \sum_i J_\perp{\bf S_\perp}_i\cdot {\bf S_\perp}_{i+1} + J_zS_{zi}S_{zi+1} + {\bf H}\cdot{\bf S}.
\end{equation}
Here ${\bf S_i}=({\bf S_{\perp i}},S_{zi})$ is a quantum spin (either integer or half-integer)
at chain site $i$.  We allow both the spin-chain to have anisotropic couplings $(J_\perp,J_z)$
and for a magnetic field, ${\bf H}$, to be potentially present.  We are interested in computing
the DSF
\begin{equation}\label{eii}
\chi(\omega,q)\!=\!-\!\int\! d\tau dx
e^{i\tilde\omega \tau-iqx}
\langle {\bf S}(\tau,x) {\bf S}(0)\rangle\bigg|_{\tilde\omega\rightarrow-i\omega+\delta}.
\end{equation}
To compute this quantity, we expand $C(\tau ,x) \equiv\langle {\bf S}(\tau ,x) {\bf S}(0)\rangle$
in a basis, $\{|l\rangle\}$, of exact eigenstates of $H$, 
\begin{equation}\label{eiii}
C(\tau ,x)=\frac{1}{\cal Z}
\sum_{l,m} e^{-\beta E_l}
\langle l|S_z(\tau ,x)|m\rangle\langle m|S_z(0)|l\rangle,
\end{equation}
where $E_l$ is the energy of eigenstate, $|l\rangle$, and ${\cal Z} = \sum_le^{-\beta E_l}$ is the
partition function of the theory.  By virtue of the gap, $\Delta$, in this system,
the Fourier transform of $C(\tau ,x)$ has a well defined low temperature expansion.  

This representation of the DSF finds its virtue when we employ a continuum,
integrable reduction of the (lattice) model in Eqn. (\ref{ei}).  In such cases the matrix
elements $\langle l|S_z(0)| m\rangle$ can readily be computed exactly.  At $T=0$ this permits
the exact computation of ${\rm Im} \chi(\omega ,q)$ at energies, $\omega$, in the vicinity of the gap through the computation
of a small number of matrix elements.  At finite temperatures, this approach, for the problem at hand, breaks down in two fashions:
i) the needed matrix elements (as well as ${\cal Z}$) become highly singular objects; and ii) to obtain the finite temperature
broadening of the coherent mode, an infinite number of matrix elements are needed.  We solve these problems
in a two step fashion.   The singularities of the matrix elements are intimately associated with treating
the spin chain as infinite in length.  While it is possible in certain circumstances to deal with these
singularities directly,\cite{smirnov,balog,muss,rmk,alt}
to circumvent this first difficulty, we instead work with chains of large but finite length, $R$.
The infinities in the matrix elements are then reduced to terms merely proportional to $R$ which are cleanly cancelled by
similar terms in the partition function.  As part of this, we will exploit the fact that the matrix
elements, even at finite $R$, are computable up to exponentially small corrections.
To handle the second difficult, we recognize that the {\it infinite} 
subset of needed matrix elements from the sum in Eqn. (\ref{eiii}) are organized
according to a Dyson's equation.  This allows us to characterize the subset by resumming a {\it finite} number of matrix elements.
We now consider how this approach works in practice in two experimentally relevant
cases, the transverse field Ising model and the spin-1 chain
as represented by the O(3) non-linear
sigma model.

\noindent{\bf Transverse field Ising model:}
The transverse field Ising model (TFIM) is obtained by taking $S_i\cdot S_i = 3/4$ in Eqn. (1),
and setting $J_\perp =0$, ${\bf H} = H\hat x$.  In the vicinity
of the TFIM's critical point (i.e. $J_z = H$), this theory has a continuum representation
as a free Majorana fermion:\cite{zamo}
\begin{equation}\label{eiv}
H = \frac{1}{2\pi}\int^R_0 dx \frac{v}{2}({\bar\psi} \partial_x\bar\psi + \psi\partial_x\psi) - i\Delta\psi{\bar\psi}.
\end{equation}
Here $\psi(x,t)$ and $\bar\psi (x,t)$ are the right and left components of a Majorana Fermi field.
The gap, $\Delta$, of the fermions in the disordered regime ($J_z > H$) 
is given by $\Delta \sim (J_z - H)$.  The Hilbert space
of the theory on a periodic line of finite length $R$ divides itself into two sectors: Neveu-Schwarz (NS) and
Ramond (R).  The NS-sector consists of a Fock space built with 
even numbers of half-integer fermionic modes, 
i.e. states of the form $|p_1\cdots p_{2N}\rangle_{NS} \equiv a^\dagger_{p_1}\cdots a^\dagger_{p_{2N}}|0\rangle_{NS}$ 
where a mode's momentum satisfies, $p_i=2\pi(n_i+1/2)/R$, with $n_i$ an integer,
while the R-sector consists of a Fock
space composed of odd numbers of even integer fermionic modes, 
$|k_1\cdots k_{2M+1}\rangle_R \equiv \{ a^\dagger_{k_1}\cdots a^\dagger_{k_{2M+1}}|0\rangle_R\}$,
$k_i=2\pi n_i/R$.  The energy/momentum, $E(p_i)/P(p_i)$, of a NS state, $|p_1\cdots p_{2N}\rangle_{NS}$,
is given simply by $E(p_i)/P(p_i) = \sum_i^{2N} \epsilon(p_i)/\sum_i^{2N}p_i$ 
where $\epsilon(p) = \sqrt{p^2+\Delta^2}$, with an identical relation holding for states in the R-sector.

To compute the DSF of this model, we need access to the matrix elements, $\langle l | S_z | m \rangle$,
of the spin operator $S_z$ at finite $R$.  These matrix elements, derived in Refs. (\onlinecite{bugrij,zamo}), only 
are non-zero when $| l \rangle$ and $| m \rangle$ belong
to different sectors.  For such matrix elements we have 
\begin{eqnarray}\label{ev}
_R\langle k_1\cdots k_{2M+1}|S_z(0)|p_1\cdots p_{2N}\rangle_{NS} \!\!&=&\!\! C_R\prod_{i,j}g(\theta_{k_i})g(\theta_{p_j})\cr
&& \hskip -2.3in \times \prod_{i<j}f(\theta_{k_i}-\theta_{k_j})\prod_{i<j}f(\theta_{p_i}-\theta_{p_j})\prod_{i,j} f^{-1}(\theta_{k_i}-\theta_{p_j}),
\end{eqnarray}
where $\theta_{p_i}$ parameterizes the momentum $p_i$ via $p_i = \Delta \sinh(\theta_{p_i})$, $C_R = 1 + {\cal O}(e^{-\Delta R})$, 
$g(\theta) = (\Delta R\cosh(\theta))^{-1/2}$, and $f(\theta) = \tanh (\theta/2)$.  Note that up to exponentially
small corrections, these matrix elements have the same functional form as at $R=\infty$.  The sole difference
in the two cases is that at finite $R$, the momenta are quantized.  This is a pattern that 
repeats itself for general integrable
models, as emphasized in Ref. \onlinecite{takacs}, and that we will exploit for our analysis of spin-1 chains.

Crucially all of these matrix elements are finite, a consequence of working at finite $R$.  The sole possible
divergence comes from the term, $\prod_{i,j} f^{-1}(\theta_{k_i}-\theta_{p_j})$, and occurs
as two momenta, $k_i$ and $p_j$, approach one
another.  But as $k_i$ lies in the R-sector with integer quantization and $p_j$ lies in the NS-sector with
half-integer quantization, the two are never exactly equal provided $R$ is finite.  In contrast, with $R=\infty$
the distinction between the R- and NS-sectors collapses (via a spontaneous $Z_2$ symmetry breaking).  
Concomitantly, $S_z$ has matrix elements where $k_i$ and $p_j$ may, in principle, be equal, and so which
are infinite.  By working at finite $R$, we thus obtain a clean, unambiguous regulation of these infinities. 

Even though any given matrix element is finite, we must still sum an infinite number of matrix
elements in the Lehmann expansion of Eqn. (\ref{eiii}) in order to obtain the DSF, $\chi^I$, at finite T.
To do so we employ a Dyson like equation by writing $\chi^{I} (\omega, q)$ in the form,
\begin{equation}\label{evi}
\chi^{I}(\omega, q) = D^{I}(\omega, q)/(1-D^{I}(\omega, q)\Sigma^{I}(\omega, q)).
\end{equation}
$D^{I}(\omega,q)$ is the DSF in the absence of temperature induced interactions:
$D^{I}(\omega,q) = 2\epsilon(q)/((\omega+i\delta)^2+\epsilon^2(q))$.  As $\chi^{I}(\omega, q)$ has
a well-defined low temperature expansion, so must $\Sigma^{I}(\omega, q)$:
%\begin{equation}\label{evii}
%\Sigma^{I}(\omega, q) = \sum_n \Sigma^{I}_n(\omega, q),
%\end{equation}
$\Sigma^{I}(\omega, q) = \sum_n \Sigma^{I}_n(\omega, q)$,
where $\Sigma^{I}_n$ is at best ${\cal O}(e^{-n\beta\Delta})$.
We can readily compute $\Sigma^{I}_1$.  To do so we expand $\chi^{I}$ to first
order in $\Sigma^{I}_1 $:
$\chi^{I}  = D^{I} + (D^{I})^2\Sigma^{I}_1 + {\cal O}(e^{-2\beta\Delta})$.
We then compare this expansion to the expansion of $\chi^{I}$ in terms of the Lehmann expansion of Eqn. (\ref{eiii}).
To facilitate this we divide $C^{I}(x,\tau)$ into contributions coming from matrix elements with a fixed number of excitations on
either side of the operator, $S_z$, i.e.
$C^{I}(x,\tau) = \sum_{M,N}C^{I}_{M,N}(x,\tau)$ where
\begin{eqnarray}\label{eix}
C^{I}_{M,N}(x,\tau) \!\!&=&\!\!\!\!\!\! \!\!\sum_{k_1,\cdots,k_{2M+1} \atop p_1,\cdots,p_{2N}}\!\!\!\!\!\!|_R\langle k_1\cdots k_{2M+1}|S_z(0)|p_1\cdots p_{2N}\rangle_{NS}|^2\cr
&& \hskip -.5in \times e^{-\beta E(k_j)}e^{-\tau(E(p_i)-E(k_j))+ix(P(p_i)-P(k_j))}.
\end{eqnarray}
Kinematic constraints give
that $C^{I}_{M,N}(\omega, q)$ is of order $e^{-\beta{\rm max}(2N\Delta-\theta(\omega)\omega,(2M+1)\Delta+\theta(-\omega)\omega)}$.
Keeping terms to at least ${\cal O}(e^{-\beta\Delta})$ and such that ${\rm Im}\chi^{I}(-\omega, q) =-{\rm Im}\chi^{I}(\omega, q)$,
we reduce $\chi^{I} (\omega, q)$ to
$\chi^{I} = \frac{1}{{\cal Z}}(C^{I}_{01}+C^{I}_{10}
+ C^{I}_{12}+ C^{I}_{21} + e^{-2\beta\epsilon(q)}D^I)$.
The final term, $e^{-2\beta\epsilon(q)}D^I$, is a `disconnected' contribution
arising from $C_{23}+C_{32}$, exactly cancelling off a similar contribution appearing in $C_{21}+C_{12}$.
Comparing these two expansions for $\chi^{I}$ gives us an expression for $\Sigma_1$.
First expanding out the partition function, ${\cal Z}^{I} = \sum_{n=0}^\infty Z^{I}_n$ where $Z^{I}_0=1$, 
$Z^{I}_1 = \sum_{p\in R}e^{-\beta\epsilon(p)}$ and generally $Z^{I}_n$ is ${\cal O}(e^{-n\beta\Delta})$,
and then noting that $C^{I}_{01}+C^{I}_{10} = (1-e^{-\beta \epsilon(q)})D^{I}$,
we obtain for $\Sigma^{I}_1$,
$\Sigma^{I}_1 
= (C^{I}_{12}+C^{I}_{21})(D^{I})^{-2}-(Z^{I}_1+e^{-\beta\epsilon(q)})(1-e^{-\beta\epsilon(q)})(D^{I})^{-1}$.
To then evaluate $\Sigma^{I}_1$, we compute $C^{I}_{12}$ and $C^{I}_{21}$ numerically.  As validation of our
use of a finite $R$ regulation of the singularities in the matrix elements, $\Sigma^{I}_1$ 
remains finite as $R\rightarrow \infty$ even though $C^{I}_{12}$, $C^{I}_{21}$, and $Z^{I}_1$
all diverge.

\begin{figure}[tbh]
\vskip 0in
\includegraphics[height=2.2in,width=2.7in,angle=0]{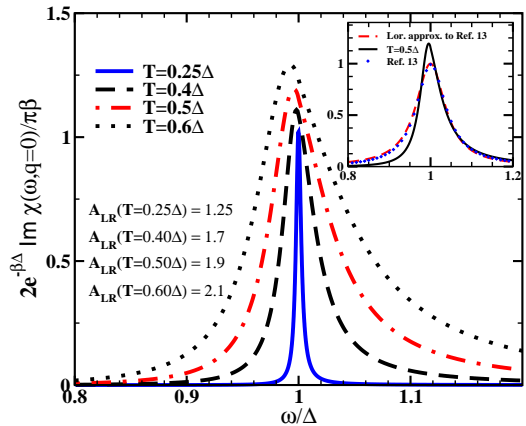}
\caption{Plots of the finite T DSF for the TFIM.}
\end{figure}

In Fig. 1 we plot the resulting DSF, ${\rm Im}\chi^{I}(\omega, q=0)$, at a variety of temperatures.  
At $T=0.25\Delta$, the DSF is approximately Lorentzian, but as the temperature is increased to $T=0.6\Delta$,
the lineshape develops a marked asymmetry.
This asymmetry was also found to be present in a virial-like expansion of the finite T DSF.\cite{reyes} 
We quantify the amount of asymmetry in the lineshape by computing the ratio, $A_{\rm LR}$, of the spectral weight to
the left and the right of $\omega = \Delta$.  In the inset of Fig. 1 we compare our results for $T=0.5\Delta$
(black solid curve) to those arrived at by using a semi-classical approach \cite{young} (blue curve)
and to a Lorentzian approximation thereof (red dashed curve).  We see that
our computation yields asymmetries in the lineshape far in excess of those found in the semi-classical approach.  
While the semi-classics has a slight asymmetry at $T=0.5\Delta$, it is close to being Lorentzian.

The origin of this discrepancy between the semi-classics and our treatment lies in two factors.  
The TFIM, as written in
Eqn. (\ref{eiv}), is relativistically invariant whereas the semi-classical model used in Ref.\onlinecite{damle}
only possesses Galilean invariance.  It has already been noted that  
a relativistic dispersion relation, in comparison with a Galilean invariant one, 
better matches the measured lineshape.\cite{kenz3}
A further difference
arises in that the semi-classics for the TFIM is only strictly correct in the $T\rightarrow 0$ limit.  At
finite $T$ it misses corrections that here are 
encoded in the form of the matrix elements (Eqn. (\ref{ev})).

\noindent{\bf Spin-1 Heisenberg model:}
We now apply our approach to the thermal broadening of the coherent mode in a gapped isotropic spin-1 chain 
(i.e. taking ${\bf S}\cdot{\bf S}=2$, $J_\perp=J_z\equiv J$, and
$H=0$ in Eqn. (\ref{ei})) .  The isotropic spin 1-chain is given in the continuum limit by
the O(3) non-linear sigma model:\cite{haldane}
\begin{equation}\label{exii}
{\cal L} = (2g)^{-1} (\partial_x{\bf n}\cdot\partial_x{\bf n}+\partial_\tau{\bf n}\cdot\partial_\tau{\bf n}).
\end{equation}
The lattice spin operators, $S_i$, are related to the continuum fields by
${\bf S}_i \simeq (-1)^{i} {\bf n}(ia_0) +
\frac{1}{g}\,{\bf n}\times\partial_t{\bf n}$ (with $a_0$ the lattice spacing).\cite{o3}  In this letter we will
focus on the DSF near wavevector $q=\pi$ and so be interested 
in computing $C(x,\tau) = \langle n^z(x,\tau)n(0)\rangle$.
The spectrum and scattering matrix of the O(3) nonlinear sigma model (NLSM)
are known exactly. There are three elementary excitations, $A^\dagger_a(\theta)$, $a=x,y,z$, forming a
vector representation of O(3).  The excitations have
a gap behaving as $\Delta \sim J e^{-1/g}$.
The excitations' energy and momentum are
parametrized in terms of the rapidity $\theta$ via
$\epsilon(\theta)=\Delta\cosh(\theta)$ and $p(\theta)=\Delta\sinh(\theta)$.

Like the TFIM, the eigenstates of the O(3) NLSM can be delineated exhaustively
in terms of multi-excitation states, i.e. $|\theta_1,a_1;\cdots,\theta_n,a_n\rangle =
A^\dagger_{a_1}(\theta_1)\cdots A^\dagger_{a_N}(\theta_N)|0\rangle$.
However, the matrix elements of these states involving the operator, $n_z$, are considerably more 
complicated than those
of the TFIM.  But as we are working at low temperatures, to compute the DSF,
$\chi^{O_3}$, we will only need recourse to matrix elements involving
a maximum of three excitations (as with the TFIM).  
In infinite volume they are given by:\cite{smirnov,bn}
\begin{eqnarray}\label{exiii}
&&\langle 0|n^a(0)|\theta,b\rangle=\delta_{ab}\ ,\cr
&&\langle \theta_1,a_1|n^a(0)|\theta_3,a_3;\theta_2,a_2\rangle
=-\frac{\pi^{3}}{2}\psi(\hat\theta_{12})\psi(\hat\theta_{13})\psi(\theta_{23})\nn
&&\times\Bigl[
\delta_{aa_1}\delta_{a_2a_3}\theta_{23}
+\delta_{aa_2}\delta_{a_1a_3}\hat\theta_{31}+\delta_{aa_3}\delta_{a_1a_2}\hat\theta_{12}
\Bigr],
\end{eqnarray}
where $\psi(\theta)=\frac{\theta+i\pi}{\theta(2\pi i+\theta)}
\frac{\tanh^2(\theta/2)}{\theta}$, $\hat\theta = \theta-i\pi$, and $\theta_{12}=\theta_1-\theta_2$.

As with the TFIM model, we work in finite volume.  The sole effect of doing so upon the matrix elements, 
up to negligible $e^{-\Delta R}$ corrections, 
is to quantize the momentum (i.e the $\theta$'s) with an attendant effect upon finite
volume phase space.\cite{takacs}  Here however the quantization
is more complex than that of the TFIM model.  We must take into account the non-trivial interactions between
the excitations and solve the Bethe ansatz equations.   For the calculation at hand, we, at most, 
most solve the one and two
particle Bethe equations for the states, $|\theta_1,a_1\rangle$ and $|\theta_3,a_3;\theta_2,a_2\rangle$.  
For the one particle state, we have the free quantization condition, $\theta_1 = \sinh^{-1}(2\pi n/R)$ for
some integer $n$.  For the two particle case, $\theta_2/\theta_3$'s are quantized via
$e^{iR\Delta\sinh(\theta_2/\theta_3)}=e^{i\delta_\alpha(\theta_{23}/\theta_{32})}$,
where the non-trivial phase, $\delta_\alpha (\theta)$, marks the presence of interactions and depends
upon the particular SU(2) representation ($\alpha =$ singlet/triplet/quintet) into which the two 
particle state falls.
Because of the presence of interactions, the finite $R$ matrix elements of the form
$\langle \theta_1,a_1|n^a(0)|\theta_3,a_3;\theta_2,a_2\rangle$ are never infinite as the $\theta$'s 
never coincide.  We again
see finite $R$ provides a clean regulation of the singularities present at $R=\infty$.

The remainder of the calculation of $\chi^{O_3}$ follows in exact analogy to the TFIM.  
$\chi^{O_3}(\omega ,q)$ takes the form of Eqn. (\ref{evi}) with 
$D^{O_3} = D^I$.  In this case the expansion of $C^{O_3}(x,\tau) \equiv \langle n^z(x,\tau)n^z(0)\rangle$,
$C^{O_3}(x,\tau) = \sum_{M,N}C^{O_3}_{M,N}(x,\tau)$, appears as
\begin{eqnarray}\label{exiv}
C^{O_3}_{M,N}(x,\tau) &=& \!\!\!\!\! \sum_{\Theta_1;\cdots;\Theta_M \atop \Theta'_1;\cdots;\Theta'_N}
|\langle \Theta_1,\cdots,\Theta_M|n^z(0)|\Theta'_1;\cdots;\Theta'_N\rangle|^2\cr
&& \hskip -.5in \times e^{-\beta E(\theta_j)}e^{-\tau(E(\theta_i)-E(\theta_j))+ix(P(\theta_i)-P(\theta_j))},
\end{eqnarray}
where $\Theta_n \equiv \{\theta_i,a_i\}$.  The partition function
here admits the expansion, $Z^{O_3} = 1+Z^{O_3}_1 + {\cal O}(e^{-2\beta\Delta})$, and 
$Z^{O_3}_1 = 3\sum_n e^{\beta\Delta\cosh(\theta_n)}$
With these redefinitions of $C_{M,N}$, $D$ and $Z_n$, $\Sigma^{0_3}_1$ takes 
the same functional form as $\Sigma^{I}_1$.

\begin{figure}[tbh]
\includegraphics[height=2.2in,width=2.8in,angle=0]{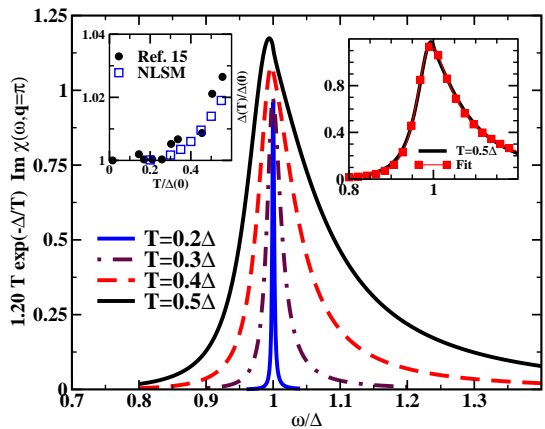}
\caption{Plots of the finite T DSF for the O(3) NLSM.}
\end{figure}

In Fig. 2 we plot the resulting finite T DSF for the O(3) NLSM.
We again find that the lineshape
is characterized by an asymmetry that grows with temperature.  We observe that this asymmetry is
far stronger than that of the semi-classical analysis (see Fig. 10 of Ref. \onlinecite{damle}) whose
lineshape is well described by a Lorentzian even up to temperatures of $T = 0.5\Delta$.  The origin
of this discrepancy between our treatment and the semi-classics 
is similar to that of the TFIM, i.e. finite energy effects in both the scattering
of excitations and in the matrix elements involving the operator, $n_z$.

The asymmetry in the lineshape can be interpreted in terms of a temperature dependent gap, $\Delta (T)$.\cite{xu}  
The gap can be extracted as the location of the peak of a Lorentzian fitted to the asymmetric lineshape.  
We plot $\Delta (T)$ vs $T$ in the left inset to Fig. 2 and compare our computations with the neutron
scattering measurements performed on the spin chain, $YBaNiO_5$, in Ref. (\onlinecite{xu}).  We see
good agreement.  For other purposes, for example, the analysis
of the three magnon scattering continuum in the spin chain compound $CsNiCl_3$
for temperatures $T>0.25\Delta$, 
we suggest the fitting function,
\begin{equation}\label{exv}
{\rm Im}\chi(\omega,q) = A/((\omega - \epsilon(q) - B)^2+C)^{1-D(x-1)}.
\end{equation}
In the right inset to Fig. 2 we show that this function provides a good fit of our computed $T=0.4\Delta$ lineshape.  

In conclusion we have presented a method by which the lineshape of the coherent mode of gapped spin chains
can be determined at finite temperature.  
This method, employing a continuum integrable representation of the chains,
works with finite length, $R$, systems so as to circumvent infinities 
in matrix elements that appear at $R=\infty$.  It further
employs a Dyson-like resummation of the matrix elements appearing in a Lehmann expansion of the DSF.
The primary conclusion drawn from this analysis is that the 
lineshape of the mode is asymmetric with an asymmetry increasing
with temperature.

We are grateful to 
A.M. Tsvelik for helpful discussions and communications and G. Xu for
access to data from Ref. 15.  This work was
supported by the EPSRC under grant GR/R83712/01 (FHLE), the DOE under
contract DE-AC02-98 CH 10886 (RMK) and the SCCS Theory Institute
at BNL (FHLE).
%We thank the ICTP Trieste, where part of this work was carried out,
%for hospitality.

%%%%%%%%%%%%%%%%%%%%%%%%%%%%%%%%%%%%%%%%%%%%%%%%%%%%%%%%%%%%%%%%%%5

\end{document}